\documentclass[letterpaper, paper,11pt]{AAS}

\usepackage{bm}
\usepackage{amsmath}
\usepackage{subfigure}
\usepackage{comment}
\usepackage{overcite}
\usepackage{footnpag}			      	
\usepackage{amssymb}
\usepackage{subfigure}   
\usepackage{multicol}
\usepackage{multirow}
\usepackage{booktabs}
\usepackage{pstool}             
\usepackage[colorlinks]{hyperref}

\usepackage{amsthm}

\PaperNumber{22-519}

\begin{document}

\title{Uncertainty Propagation and Filtering via the Koopman Operator in Astrodynamics}

\author{Simone Servadio\thanks{Postdoctoral Associate, Department of Aeronautics and Astronautics, Massachusetts Institute of Technology, MA 02139, USA. email: simoserv@mit.edu},
\ William Parker\thanks{SM Candidate, Department of Aeronautics and Astronautics, Massachusetts Institute of Technology, MA 02139, USA. email: wparker@mit.edu}, 
\ and Richard Linares\thanks{Boeing Assistant Professor, Department of Aeronautics and Astronautics, Massachusetts Institute of Technology, MA 02139, USA. email: linaresr@mit.edu}
}

 \maketitle
 
\begin{abstract}
The Koopman Operator (KO) provides an analytical solution of dynamical systems in terms of orthogonal polynomials. This work exploits this representation to include the propagation of uncertainties, where the polynomials are modified to work with stochastic variables. Thus, a new uncertainty quantification technique is proposed, where the KO solution is expanded to include the prediction of central moments, up to an arbitrary order. The propagation of uncertainties is then expanded to develop a new filtering algorithm, where measurements are considered as additional observables in the KO mathematics. Numerical simulations in astrodynamics assess the accuracy and performance of the new methodologies.

\end{abstract}
 
 
\section{Introduction}

The development of capable general methods for propagating the orbital motion of a satellite subject to strongly nonlinear dynamics is an important challenge in modern astrodynamics. Along with predicted orbital motion itself, an awareness of the accumulated uncertainty in the prediction is useful for a wide variety of operational applications relating to analysis, estimation, and control tasks. Uncertainty propagation is usually followed by the measurement update, where the system receives noisy observations from the sensors, and corrects its predicted state into its estimate [\citen{bar2004estimation}]. The final result constitutes a complete filtering algorithm. For the linear and Gaussian case, the Kalman formulation [\citen{kalman1960new,kalman1961new}] represents the optimal solution, since the conditional probability density function (PDF) remains Gaussian at all times and it is evaluated through Bayes's rule [\citen{simon2006optimal}].  

However, most of the systems in astrodynamics are affected by nonlinearities, which leads to a sub-optimal solution for the evaluation of the conditional PDF [\citen{schutz2004statistical}]. The first solution to address nonlinearities is the Extended Kalman Filter (EKF) [\citen{gelb1974applied}], where the estimation error is linearized around the most current estimate and the Kalman formulation is applied. However, it has been shown that the linearization assumption is insufficient to correctly propagate uncertainties for orbit determination applications [\citen{junkins2004nonlinear}]. A more robust technique is offered by the Unscented Transformation (UT) [\citen{julier2000new}], where carefully selected sigma points undergo the full nonlinear dynamics to propagate central moments as weighted means. The Unscented Kalman Filter (UKF) [\citen{julier2004unscented}] applies UT for state and measurement prediction, obtaining superior performance than the EKF. 

The propagation of uncertainties in orbital mechanics has been widely addressed. Park and Scheeres [\citen{park2006nonlinear}] propagate mean and high order central moments using state transition tensors (STT) through arbitrary nonlinear dynamics. They, later [\citen{park2007nonlinear}], complete a filtering algorithm able to include measurement update. Majji, Turner and Junkins [\citen{majji2008high}] expand this work to update all the higher central moments. Afterwards, Valli \textit{et al.} are able to recreate these results using differential algebra (DA) techniques [\citen{valli2013nonlinear}], with applications to orbit determination and rendezvous maneuvers [\citen{servadio2021nonlinear,cavenago2018based}].  

This work proposes a novel uncertainty propagation technique, based on the Koopman Operator (KO). KO enables the re-formulation of strongly nonlinear dynamics into a linear polynomial framework. The KO theory was initially developed by Bernard Koopman and further developed by John Von Neumann in the 1930s [\citenum{koopman1931hamiltonian,neumann1932operatorenmethode}]. Koopman and Von Neumann applied the operator theoretic perspective used in the development of quantum mechanics [\citenum{naylor2000,prigogine}] to classical mechanics problems. They introduced the Koopman Operator: a linear operator that can reformulate a nonlinear dynamical system of finite dimension into a linear system with an infinite number of dimensions. To leverage the benefit of working with linear systems, a finite-dimensional approximation of the dynamics can be attained by observing the projection on a finite subspace spanned by a set of basis functions. Recent interest in studying the evolution of nonlinear dynamical systems in fluid mechanics [\citenum{mezic2013analysis}], control [\citenum{brunton2016koopman,surana2016koopman}], and attitude dynamics [\citenum{chen2020koopman}] has led to a fruitful re-discovery of the KO theory for spectral analysis and linear approximations of these dynamical systems. 

More recently, KO has been employed to address computational challenges in astrodynamics with promising results. The operator theoretic framework was first introduced to model satellite motion in the proximity of libration points [\citenum{libration}]. Koopman-based perturbation theory has been used to model the motion of a satellite about an oblate planet [\citenum{arnas2021analysis, arnas2021koopman}] and an approximate analytical solution to the zonal harmonics problem was identified [\citenum{arnas2021approximate}]. In this work, KO approximates PDFs via their central moments, which are propagated using the KO eigendecompostion of the system with orthogonal polynomials. The KO analytical solution of the system provides a global analysis of the dynamics, such that it can overcome the drawbacks shown by local approximations, such as DA and the Taylor polynomial solutions, where large uncertainties need to be split,    using Automatic Domain Splitting (ADS) [\citen{wittig2015propagation}] or Gaussian Multiple Models (GMM) [\citen{servadio2021differential}], to maintain accurate performance. 

The paper is divided as follows. The KO theory is presented in the next section. Afterwards, the new uncertainty propagation technique and the filtering algorithm are presented. Then, uncertainty quantification methodology is applied to the complex dynamics of satellite motion under the circular restricted three body problem (CRTBP) [\citen{servadio2021dynamics}], with applications to orbit determination with limited measurements. Lastly, conclusions are drawn.


\section{Koopman Operator Theory}
A nonlinear dynamical system can be expressed by the initial value problem, which introduces a set of coupled autonomous ordinary differential equations (ODEs) of the form
\begin{equation}
\left\{ \begin{tabular}{l}
    $\displaystyle\frac{d}{dt}{\bf x}(t) = {\bf f}({\bf x})$ \\
    ${\bf x}(t_0) = {\bf x_0}$ 
\end{tabular}  \right.,
\end{equation}
where ${\bf x}\in\mathbb{R}^d$ is the state  which depends on the time evolution $t$, ${\bf f}: \mathbb{R}^d\rightarrow \mathbb{R}^d$ is the nonlinear dynamics model, $d$ is the number of dimensions in which the problem is defined, and ${\bf x_0}$ is the initial condition of the system at time $t_0$. The Koopman Operator $(\mathcal{K})$, propagates $\mathcal{G}({\bf x})$, the functions describing observables of the state. 

The KO is an infinite-dimensional operator. A result of this property is that the vector space of the observable functions, defined here as $\mathcal{F}$ (where $\mathcal{G}({\bf x}) \in \mathcal{F}$) is also infinite-dimensional.  It follows that if $g\subseteq\mathcal{G}({\bf x})$ is a given observable in this space, the evolution of $g$ in the dynamical system can be found by
\begin{equation}
\mathcal{K}\left(g({\bf x})\right) = \frac{d}{dt}g({\bf x}) = \left( \nabla_{{\bf x}} g({\bf x})\right)\frac{d}{dt}{\bf x}(t) = \left( \nabla_{{\bf x}} g({\bf x})\right){\bf f}({\bf x}),
\end{equation}
where $\nabla_{{\bf x}} g = (\partial g/\partial x_1,\partial g/\partial x_2,\dots,\partial g/\partial x_d)$. More generally, any observable governed by the dynamical system can be evolved using the KO in this way; 
\begin{equation}
\mathcal{K}\left(\cdot\right) = \left( \nabla_{{\bf x}} \cdot\right){\bf f}({\bf x}).
\end{equation}
Koopman's initial work [\citenum{koopman1931hamiltonian}] outlined the linearity property of the KO, which implies 
\begin{equation}
    \mathcal{K}\left(\beta_1g_1({\bf x})+\beta_2g_2({\bf x})\right)=\beta_1\mathcal{K}\left(g_1({\bf x})\right)+\beta_2\mathcal{K}\left(g_2({\bf x})\right),
\end{equation}
for any pair of observables $g_1\subseteq\mathcal{G}({\bf x})$ and $g_2\subseteq\mathcal{G}({\bf x})$ and any arbitrary constants $\beta_1$ and $\beta_2$. 
The linearity of the KO allows for efficient computation, but the infinite-dimensionality of the operator makes it impossible to capture the evolution of all measurement functions in a Hilbert space [\citen{brunton2016koopman}].  This issue can be mitigated, however, by instead capturing the evolution of the system on a finite subspace spanned by a finite set of basis functions. Performing this step truncates the KO to a finite subspace $\mathcal{F}_D$ of dimension $m$, where $\mathcal{F}_D \in \mathcal{F}$. This subspace $\mathcal{F}_D$ can be spanned by any set of eigenfunctions $\phi_i\in\mathcal{F}_D$, with $i\in\{1,2,\dots,m\}$, defined as
\begin{equation}\label{koopman}
\mathcal{K}\left(\phi_i({\bf x})\right) =\frac{d}{dt}\phi_i({\bf x})=\lambda_i \phi_i({\bf x}),
\end{equation}
where $\lambda_i$ are the eigenvalues associated with the eigenfunctions $\phi_i$, and $m$ is the number of eigenfunctions chosen to represent the space. To linearize the system, a transformation of variables is performed using the Koopman eigenfunctions, ${\bf \Phi}({\bf x}) = \left(\phi_1({\bf x}), \dots, \phi_m({\bf x}) \right)^T \in \mathcal{F}_D$. Then, using the relation in Eq.~\eqref{koopman}, the evolution of ${\bf \Phi}$ can be expressed as 
\begin{equation}\label{time_eigenf}
\mathcal{K}\left(\bf \Phi\right) = \frac{d}{dt}{\bf \Phi}=\Lambda {\bf \Phi},
\end{equation}
where $\Lambda=\text{diag}([\lambda_1, \dots,\lambda_m])$ is the diagonal matrix containing the eigenvalues of the system in $\mathcal{F}_D$.  The solution of Eq.~\eqref{time_eigenf} is
\begin{equation}\label{eigen_time}
    {\bf \Phi}(t) = \exp(\Lambda t){\bf \Phi}(t_0),
\end{equation}
where ${\bf \Phi}(t_0)$ is the value of the eigenfunctions at the initial time $t_0$. Once the KO eigenfunctions are computed, Eq. ~\eqref{eigen_time} evolves them in time. For the general case, there is interest in the identity observable, that is, ${\bf g}({\bf x})={\bf x}$.

\subsection{Computing the Koopman Eigenfunctions via Galerkin Method}

Following the theory developed in [\citenum{arnas2021analysis}], we noticed that the KO gives rise to a set of linear Partial Differential Equations (PDE) for the eigenfunctions:
\begin{equation}\label{koopman2}
\mathcal{K}(\phi_i) = \left(\nabla_x \phi_i({\bf x})\right) {\bf f}({\bf x}) =\lambda_i .
\end{equation}
The Galerkin method offers an approximation of the solution of this PDE. Let us select the Legendre polynomials as the set of basis functions, since they create an orthogonal basis and have computational advantages in the computation of the Koopman matrix [\citen{zonal_koopman}]. The main idea of the methodology is to represent any function of the space as a linear combination of the basis functions. This is achieved by defining an inner product:
\begin{equation}
\langle  f, g \rangle =\int_{\Omega} f({\bf x})g({\bf x}) w({\bf x})d{\bf x},
\end{equation}
where $f$ and $g$ be two arbitrary functions, and $w({\bf x})$ is a positive weighting function defined on the space domain $\Omega$. For the case of Legendre polynomials, the weighting function is a constant $w({\bf x}) = 1$, and the domain for each variable ranges between $[-1,1]$. In addition, the normalized Legendre polynomials are defined such that:
\begin{equation}
\langle  L_i, L_j \rangle =\int_{\Omega} L_i({\bf x})L_j({\bf x}) w({\bf x})d{\bf x} = \delta_{ij},
\end{equation}
where $L_i$ and $L_j$ with $\{i,j\}\in\{1,\dots,m\}$ are two given normalized Legendre polynomials from the set of basis functions selected, and $\delta_{ij}$ is Kronecker's delta. By using this set of orthogonal multivariate polynomials, the KO eigenfunctions can be represented as a series expansion in terms of this set of basis functions, particularly:
\begin{equation}\label{series}
\phi_i({\bf x})=\sum_{\ell=1}^{\infty}c_{i\ell}L_{\ell}({\bf x})\approx \sum_{\ell=1}^{m}c_{i\ell}L_{\ell}({\bf x})={\bf c}_i{\bf L}^T({\bf x}) = {\bf L}({\bf x}){\bf c}_i^T,
\end{equation}
where $c_{i\ell}$ is a constant coefficient associated with the eigenfunction $\phi_i$ and the basis $L_{\ell}$, and ${\bf c}_i$ and ${\bf L}$ are two row vectors containing the set of coefficients $c_{i\ell}$ and the whole set of basis functions respectively. Note that although the series is infinite, a truncation was performed using $m$ different basis functions, and thus, this represents an approximation of the eigenfunctions. 

The PDE for the computation of the Koopman eigenfunction, Eq. \eqref{koopman2}, can be approximated using a Galerkin method and the series expansion from Eq.~\eqref{series}, where each function that undergoes the KO is projected onto the $m$-dimensional subspace $\mathcal{F}_D$. The time evolution of each $i$-th eigenfunction can be grouped into a diagonal system, with dimensions $m$. After some mathematical manipulations described in [\citen{arnas2021analysis}], [\citen{zonal_koopman}], and [\citen{servadio2021differential}], the diagonalization of the system is obtained as:
\begin{equation} \label{decomposition}
    C K = \Lambda C,
\end{equation}
where $C$ is the matrix containing all the coefficients $c_{ij}$, $\Lambda$ is the diagonal matrix of eigenvalues $\lambda_i$, and $K$ is the Koopman matrix, a $m \times m$ sized matrix whose components are:
\begin{equation} \label{kmatrix}
    K_{ij} = \langle \nabla_{\bf x} L_i({\bf x}) {\bf f}({\bf x}), L_j({\bf x}) \rangle.
\end{equation}
Note that Eq.~\eqref{decomposition} is in fact defining a eigendecomposition of matrix $K$, where $C$ is the matrix containing the left eigenvectors of $K$, and $\Lambda$ the associated eigenvalues. Therefore, it is possible to first obtain $K$ in closed form solution from Eq~\eqref{kmatrix}, and then perform the eigendecomposition to obtain $C$, and thus, the approximated eigenfunctions $\phi_i$ of the system through Eq.~\eqref{series}.

\subsection{Koopman modes}
The Koopman modes are the representation of the observables into the set of eigenfunctions ${\bf \Phi}$ of the system. Thus, let ${\bf g}({\bf x})$ be the observables of interest, function of the state. Using the Galerkin method, these observables can be approximated as an expansion in the set of eigenfunctions
\begin{equation}
    g_i({\bf x}) \approx \sum_{\ell=1}^m b_{i\ell}\phi_{\ell}({\bf x}),
\end{equation}
where $b_{i\ell}$ are the coefficients, computed by projecting the $i$-th observable onto the $\ell$-th basis functions:
\begin{equation}
    b_{i\ell} = \left\langle g_i, \phi_{\ell} \right\rangle.
\end{equation}
Therefore, the observables can be represented in matrix notation as ${\bf g}({\bf x}) = B {\bf \Phi}({\bf x})$, where $B$ is a matrix of size $q \times m$ containing the coefficients $b_{i\ell}$, and $q$ represents the number of observables. The time evolution of ${\bf g}({\bf x})$ is expressed by the time evolution of the eigenfunctions, or, equivalently, in terms of the basis functions:
\begin{equation}
    {\bf g}({\bf x}(t)) = B {\bf \Phi}({\bf x}(t)) = B \exp(\Lambda t){\bf \Phi}({\bf x}(t_0)) = B \exp(\Lambda t) C {\bf L}({\bf x}(t_0)),
\end{equation}
where ${\bf x}(t_0)$ are the initial condition values of the states. 

Note that the observables can be also projected onto the set of basis functions ${\bf L}$:
\begin{equation}
    g_i({\bf x}) \approx \sum_{\ell=1}^m a_{i\ell}L_{\ell}({\bf x}) \rightarrow {\bf g}({\bf x}) = A {\bf L}({\bf x})   \quad \text{where} \quad a_{i\ell} = \left\langle g_i, L_{\ell} \right\rangle,
\end{equation}
where $A_{ij}=a_{ij}$, whose solution under the system of differential equations is
\begin{equation}
    {\bf g}({\bf x}(t)) = A {\bf L}({\bf x}(t)) = A C^{-1} \exp(\Lambda t) C {\bf L}({\bf x}(t_0)).
\end{equation}

For a more detailed analysis of the analytical KO solution with orthogonal polynomials, with a step-by-step explanation of the methodology, the reader can refer to [\citen{servadio2021koopman,arnas2021analysis,zonal_koopman,servadio2021dynamics}].


\section{Uncertainty Propagation with the Koopman Operator}
The solution of the system can be formulated using the identity observable (the observable g({\bf x})={\bf x}) and the KO eigendecomposition as in the following expression:
\begin{equation} 
    {\bf g}({\bf x}(t_0), t) = A C^{-1} \exp(\Lambda t) C {\bf L}({\bf x}(t_0))
\end{equation}
Note that the solution is a function of the initial state vector at time, $t_0$, and the current time, $t$. When propagating the state up to the final time $t_f$, the Koopman solution does not require any numerical integration, and the final state ${\bf x}_f = {\bf x}(t_f)$, is directly connected to the value of the state at the initial time ${\bf x}_0 = {\bf x}(t_0)$, 
\begin{equation} \label{eq:pol}
    {\bf x}_f({\bf x}_0) = A C^{-1} \exp(\Lambda t_f) C {\bf L}({\bf x}_0).
\end{equation}
which gives the polynomial dependence of the propagated states on the initial condition. Equation \eqref{eq:pol} is a polynomial in the variable ${\bf x_0}$ and it constitutes the forward map in time of the dynamics of the system. 

When propagating PDFs, it is more convenient to work with the deviations from the mean than with the state itself as the variable [\citen{servadio2020recursive}]. This way, the central moments are easier to calculate and provide a more intuitive understanding of the uncertainties of the system when compared to raw moments [\citen{servadio2020nonlinear}]. Let us assume $\hat {\bf x}_0$ is the known mean of the initial PDF that will be propagated, in terms of its central moments, to time $t_f$. It is possible to express the Koopman solution of the state as a function of the deviation vector $\delta {\bf{x}}_0 $ from the initial condition through the variable shift ${\bf{x}}_0 = \hat {\bf x}_0 + \delta {\bf{x}}_0$:
\begin{align} 
    \tilde {\bf x}_f(\delta {\bf{x}}_0) &= A C^{-1} \exp(\Lambda t_f) C {\bf L}(\hat {\bf x}_0 + \delta {\bf x}_0) \\
    &= A C^{-1} \exp(\Lambda t_f) C \tilde{\bf L}(\delta {\bf x}_0)
\end{align}
This state polynomial represents the same map as Equation \eqref{eq:pol}, but with the variable centered at the current mean. Thus, the Legendre polynomials ${\bf L}$ and $\tilde {\bf L}$ describe the solution, but their coefficients are different such that their variable dependency has been shifted by a value $\hat {\bf x}_0$. It can be observed that the equality $ {\bf L}(\hat {\bf x}_0)= \tilde{\bf L}({\bf 0})$ holds. 

The deviations are determining all the possible outcomes of the propagated state around the center $\bar {\bf x}_f = \tilde {\bf x}_f({\bf 0} ) = {\bf x}_f(\hat {\bf x}_0 )$, which is the time propagation of the initial mean $\hat {\bf x}_0 $. In the particular case of linear dynamics and Gaussian distributions, the mean of the propagated PDF, $\hat {\bf x}_f$, coincides with the propagation of the initial mean and center of the map: $\hat {\bf x}_f = \bar {\bf x}_f$. However, due to non-linearities, the polynomial map that connects deviations from $t_0$ to $t_f$ loses its Gaussian property, and the propagated central moments need to be evaluated using expectations.

Let us assume an initial Gaussian PDF of the state, perfectly described by its mean $\hat {\bf x}_0 $ and covariance ${\bf P}_0$:
\begin{equation} 
    {\bf x}_0 \sim \mathcal{N}(\hat {\bf x}_0 ,{\bf P}_0)
\end{equation}
Thanks to the Isserlis formulation [\citen{isserlis1918formula}], it is possible to evaluate all central moments of the Gaussian distribution up to order $\psi$. The mean of the state PDF at time $t_f$ is calculated by applying the expected value operator $\mathbb E [\cdot]$ on the polynomial $\tilde {\bf x}_f(\delta {\bf{x}}_0)$ 
\begin{equation} 
    \hat {\bf x}_f = \mathbb{ E} [\tilde{\bf{ x}}_f (\delta {\bf{x}}_0)] = \bar {\bf x}_f + \sum \chi_i \mathbb{ E} [\delta {\bf x}_{0,i}]
\end{equation}
where $\chi_i$ indicates the coefficient of the $i$-th monomial in the state polynomial, and $\delta {\bf x}_{0,i}$ the relative deviation. Through linear operator property, the evaluation of the mean works directly on each single monomial of the polynomial. The expected value is evaluated by substituting, for each expectation of deviations, the relative central moment previously obtained through the Isserlis formulation [\citen{isserlis1918formula}]. Thus, having the state polynomial as a function of the deviations from the mean leads to a trivial evaluation of the expected values, where different powers of the deviations are easily connected to the relative central moments of the distribution. 

The same procedure can be adopted to evaluate central moments of any order. Starting from the propagated covariance, the expectation is given by:
\begin{equation} 
    {\bf P}_f = \mathbb{ E} [(\tilde{\bf{ x}}_f (\delta {\bf{x}}_0) - \hat {\bf x}_f)(\tilde{\bf{ x}}_f (\delta {\bf{x}}_0) - \hat {\bf x}_f)^T] 
\end{equation}
The argument of the expectation is a polynomial, where deviations get, once again, substituted by the initial central moments of the original state PDF, approximated as Gaussian. Therefore, it is possible to approximate the propagated central moments up to any arbitrary order. Let us indicate with $\tilde{\bf{ x}}_{f,i}$ the $i$-th component of the state polynomial vector; then, each component of the propagated state skewness ${\bf S}_f$ (the third-order central moment) and kurtosis ${\bf \Sigma}_f$ (fourth-order central moment) can be approximated as:
\begin{align} 
    {\bf S}_{f,ijl} &= \mathbb{ E} [(\tilde{\bf{ x}}_{f,i} (\delta {\bf{x}}_0) - \hat {\bf x}_{f,i})  (\tilde{\bf{ x}}_{f,j} (\delta {\bf{x}}_0) - \hat {\bf x}_{f,j})  (\tilde{\bf{ x}}_{f,l} (\delta {\bf{x}}_0) - \hat {\bf x}_{f,l})]\\
    {\bf \Sigma}_{f,ijlm} &= \mathbb{ E} [(\tilde{\bf{ x}}_{f,i} (\delta {\bf{x}}_0) - \hat {\bf x}_{f,i})  (\tilde{\bf{ x}}_{f,j} (\delta {\bf{x}}_0) - \hat {\bf x}_{f,j})  (\tilde{\bf{ x}}_{f,l} (\delta {\bf{x}}_0) - \hat {\bf x}_{f,l})  (\tilde{\bf{ x}}_{f,m} (\delta {\bf{x}}_0) - \hat {\bf x}_{f,m})]
\end{align}
Therefore, the state PDF is propagated in time in terms of its central moments, evaluated using polynomial substitution for each state component. 

It is worth noticing that performing the propagation of moments does not require any numerical integration. The dynamics have been solved and represented as a linear combination of well-defined eigenfunctions through the KO solution, which are connected to the basis polynomials. The propagated states, and relative moments, are evaluated directly from the Legendre polynomials once the final time is selected. This eliminates the computational expensive propagation required by other methods. 

It is possible now to complete a filtering algorithm by adding a linear update (or polynomial) to the propagated state probability density function according to the Kalman formulation [\citen{kalman1960new,kalman1961new}].


\section{The Koopman Operator Filter }
The propagation of uncertainties can be completed into a full filtering algorithm by adding the measurement update. As such, the propagation of the state mean and covariance can be discretized for the prediction of the state distribution from time step $\{t_k,{\bf{x}}_k,{\bf P}_k\}$ to time step $\{t_{k+1},{\bf{x}}_{k+1},{\bf P}_{k+1}\}$, according to the propagation equations described beforehand:
\begin{align} 
    \tilde {\bf x}_{k+1}(\delta {\bf{x}}_k) &= A C^{-1} \exp(\Lambda t_{k+1}) C {\bf L}(\hat {\bf x}_k + \delta {\bf x}_k) =
     A C^{-1} \exp(\Lambda t_{k+1}) C \tilde{\bf L}(\delta {\bf x}_k)\\
    \hat {\bf x}_{k+1}^- &= \mathbb{ E} [\tilde{\bf{ x}}_{k+1} (\delta {\bf{x}}_k)] \\
    {\bf P}_{k+1}^- &= \mathbb{ E} [(\tilde{\bf{ x}}_{k+1} (\delta {\bf{x}}_k) - \hat {\bf x}_{k+1}^-)(\tilde{\bf{ x}}_{k+1} (\delta {\bf{x}}_k) - \hat {\bf x}_{k+1}^-)^T] 
\end{align}
where $\hat {\bf x}_{k+1}^-$ and ${\bf P}_{k+1}^-$ represent the predicted mean and covariance, respectively. The state polynomial at time step $t_{k+1}$ is evaluated efficiently using the eigendecomposition of the dynamics of the system performed previously, (offline), during the Koopman spectral analysis. Therefore, the state polynomials of deviations of each state component at any time step are obtained directly from the basis functions of the system. 

The correction step (or measurement update) takes its name from modifying the predicted state mean and covariance, $\hat {\bf x}_{k+1}^-$ and ${\bf P}_{k+1}^-$, into their updated values, $\hat {\bf x}_{k+1}^+$ and ${\bf P}_{k+1}^+$. The filter receives information about the state of the system through a series of observations ${\bf y}_{k+1}$, which relationship with the state itself is known;
\begin{equation} 
    {\bf y}_{k+1} = {\bf h}({\bf x}_{k+1}) + {\boldsymbol \eta}_{k+1}
\end{equation}
where ${\bf h}(\cdot)$ is the measurement function and ${\boldsymbol \eta}_{k+1}$ is the measurement noise, with known covariance matrix ${\bf R}_{k+1}$. The aim of the algorithm is to best merge information from the prediction step and the observations to obtain the most accurate estimate of the system by ``\textit{filtering out}" the noise. 

The Koopman Operator Filter (KOF) develops the measurement update with the same polynomial algebra applied for the propagation of the state uncertainties. After all, the measurements ${\bf y}_{k+1}$ are just but a more complicated set of observables that can be represented by the set of basis functions ${\bf L}$, or $\tilde {\bf L}$. In an analogous way matrix $A$ was calculated for the identity observables to get the state of the system, a new matrix $A_y$ is evaluated to express in a polynomial way the measurement function, as a linear combination of the basis functions: ${\bf y}({\bf x}) = A_y {\bf L}({\bf x})$. Each entry of $A_y$ is evaluated using the Galerkin method with the inner product between functions, likewise any other Koopman matrix presented in the paper. After shifting the center of the polynomials using $\tilde {\bf L}$ over ${\bf L}$, the final result is the measurement polynomials as a function of the state deviation vector
\begin{align} 
    \tilde {\bf y}_{k+1}(\delta \bf{x}_k) &= A_y C^{-1} \exp(\Lambda t_{k+1}) C {\bf L}(\hat {\bf x}_k^+ + \delta {\bf x}_k) =
     A_y C^{-1} \exp(\Lambda t_{k+1}) C \tilde{\bf L}(\delta {\bf x}_k)
\end{align}
These polynomials express the distribution of the measurements as a function of the deviations of the state from the original time step $t_k$, therefore, they represent the Koopman polynomial map of deviations from the state space to the measurement space. This map directly connects to any possible state deviation, the relative outcome in the measurement space: the map tells how the measurement PDF is connected, and shaped, with respect to the state PDF. 

In order to apply the Galerkin method, and evaluate the inner product between the basis functions and the measurement observables, the measurement equations need to be expressed as a polynomial approximation of ${\bf h}(\cdot)$. For small deviations, a Taylor approximation furnish a very accurate representation of the measurement map around the current center. However, any other polynomial representation is a valid solution and its selection depends on the application. 

Following the Kalman update formulation of an estimator with a linear dependence on the measurements, the Koopman Operator Filter evaluates the predicted measurement mean by applying the expectation operator directly on the measurement polynomials
\begin{align}
    \hat {\bf y}_{k+1} &= \mathbb{ E} [\tilde{\bf{ y}}_{k+1} (\delta {\bf{x}}_k)]
\end{align}
where, once again, the mean of each component is calculated by substituting the state central moments, evaluated with Isserlis, on each monomial of the series. Continuing the analogy with the prediction step, the measurement covariance is evaluated as
\begin{align}
   {\bf P}_{yy} &= \mathbb{ E} [(\tilde{\bf{ y}}_{k+1} (\delta {\bf{x}}_k) - \hat {\bf y}_{k+1})(\tilde{\bf{ y}}_{k+1} (\delta {\bf{x}}_k) - \hat {\bf y}_{k+1})^T] + {\bf R}_{k+1}
\end{align}
having considered the influence of the measurement noise. The state-measurement cross-covariance is evaluated as
\begin{align}
   {\bf P}_{xy} &= \mathbb{ E} [(\tilde{\bf{ x}}_{k+1} (\delta {\bf{x}}_k) - \hat {\bf x}_{k+1}^-)(\tilde{\bf{ y}}_{k+1} (\delta {\bf{x}}_k) - \hat {\bf y}_{k+1})^T] 
\end{align}
so that the Kalman gain is
\begin{align}
   G &=  {\bf P}_{xy}{\bf P}_{yy}^{-1}
\end{align}
as it has been proven by the orthogonality principle [\citenum{servadio2020recursive}]. The update of the state estimate and the relative covariance is performed using the Kalman update equations
\begin{align}
   \hat {\bf x}_{k+1}^+ &= \hat {\bf x}_{k+1}^-  + G ({\bf y}_{k+1} - \hat {\bf y}_{k+1}) \\
   {\bf P}_{k+1}^+ &= {\bf P}_{k+1}^- - G {\bf P}_{yy} G^T
\end{align}
The filtering algorithm has ended and it is ready to start the next iteration, having approximated the updated state distribution as a Gaussian PDF with mean the estimate $\hat {\bf x}_{k+1}^+$ and covariance ${\bf P}_{k+1}^+$. The following step will start by evaluating the state polynomial at the new time step and shifting the basis functions at the current estimate. The expectations will be evaluated by applying the Isserlis formulation with the updated covariance matrix. 

The polynomial nature of the KOF makes it suitable for the implementation to nonlinear estimator, where the updated estimate is not a linear function of the measurement, but it has a high-order polynomial dependence on the observation that increases accuracy [\citenum{servadio2020nonlinear,servadioTaylorPoly}].

\section{The Circular Restricted Three Body Problem}
The circular restricted three body problem (CRTBP) equations can be expressed as:
\begin{subequations}\label{sys1}
\begin{align} 
\ddot{x}&=\frac{\partial \Omega}{\partial x}+2\dot{y},\\
\ddot{y}&=\frac{\partial\Omega }{\partial y}-2\dot{x},\\
\ddot{z}&=\frac{\partial \Omega }{\partial z},
\end{align}
\end{subequations}
where $\Omega=1/2\left(x^2+y^2\right)+(1-\mu)/r_1+\mu/r_2$, $\mu$ is the ratio between the mass of the secondary over the total mass of the system, and $r_1$ and $r_2$ are the distances to the primary and secondary masses. The distance $\gamma$ from the libration points, $\mathcal L_j$, to the primary body is found by solving the Euler quintic equation:
\begin{align}
\gamma_j^5\mp \left(3-\mu\right)\gamma_j^4+\left(3-2\mu\right)\gamma_j^3-\mu\gamma_j^2\pm2\mu\gamma_j-\mu=0, \, j=1,2
\end{align}

Following [\citenum{richardson1980analytic}], the equations of motion for a satellite moving near $\mathcal L_1$ and $\mathcal L_2$ can be expressed by translating the origin to the equilibrium point:
\begin{equation}
\bar{x}=\frac{x-1+\mu\pm\gamma}{\gamma},\quad \quad
\bar{y}=\frac{y}{\gamma},\quad \quad
\bar{z}=\frac{z}{\gamma},
\end{equation}
where, in the $\pm$ symbol, the upper sign relates to the dynamic about $\mathcal L_1$, while the lower sign is for $\mathcal L_2$. The CRTBP position variables $\{x,y,z\}$ have been shifted and scaled to a new set of coordinates $\{\bar x,\bar y, \bar z\}$, where the new origin is either $\mathcal L_1$ or $\mathcal L_2$, and they are normalized such that the distance between $\mathcal L_1$ or $\mathcal L_2$ and the primary is 1. For convenience of notation, $x$, $y$, $z$ will refer to $\bar{x}$, $\bar{y}$, and $\bar{z}$ variables henceforth. 

Richardson [\citenum{richardson1980analytic}] showed that the full nonlinear CRTBP equations can be expanded in terms of polynomials, especially Legendre polynomials, $P_n$, for their computational advantages. Following the developments of [\citenum{richardson1980analytic}], the CRTBP around $\mathcal L_1$ and $\mathcal L_2$ can be represented as: 
\begin{subequations}\label{CRTBP}
\begin{align}
\ddot{x}-2\dot{y}-(1+2c_2)x&=\frac{\partial}{\partial x}\sum_{n\geqslant 3} c_n \rho^n P_n\left(\frac{x}{\rho}\right),\\
\ddot{y}+2\dot{x}+(c_2-1)y&=\frac{\partial}{\partial y}\sum_{n\geqslant 3} c_n \rho^n P_n\left(\frac{x}{\rho}\right),\\
\ddot{x}+c_2z&=\frac{\partial}{\partial z}\sum_{n\geqslant 3} c_n \rho^n P_n\left(\frac{x}{\rho}\right),
\end{align}
\end{subequations}
where $\rho^2=x^2+y^2+z^2$ and the coefficients $c_n$ are given by
\begin{equation}\label{c_equation}
c_n=\frac{1}{\gamma^3}\left((\pm)^n\mu+(-1)^n\frac{(1-\mu)\gamma^{n+1}}{(1\mp\gamma)^{n+1}}\right),
\end{equation}
with upper sign for $\mathcal L_1$ and the lower one for $\mathcal L_2$. This expression can be expressed using an elegant formulation [\citenum{koon2000dynamical}], that uses the recursive property of creating Legendre polynomials. Considering the function
\begin{equation}
T_n(x,y,z) =\rho^n P_n\left(\frac{x}{\rho}\right),
\end{equation}
where $T_n$ is a homogeneous polynomial of degree $n$ in variable $x$, $y$, and $z$, and using the Legendre polynomial three-term recursive formula, it can be shown that:
\begin{equation}
T_n =\frac{2n-1}{n}xT_{n-1}-\frac{n-1}{n}\left(x^2+y^2+z^2\right)T_{n-2},
\end{equation}
where the recursion starts with $T_0 = 1$ and $T_1 = x$. This work uses this formula to define the polynomial representation of the CRTBP model.

\subsection{Hamiltonian in Normal Form}
Jorba and Masdemont [\citenum{jorba1999dynamics}] provide a summary for the evaluation of the normal form of the CRTBP Hamiltonian. The equations of motion \eqref{CRTBP} come from applying Hamilton's equation to the Hamiltonian 
\begin{equation}\label{H1}
H=\frac{1}{2}\left(p_x^2+p_y^2+p_z^2\right)+yp_x-xp_y-\sum_{n\geq2} c_n \rho^n P_n\left(\frac{x}{\rho}\right),
\end{equation}
where the variables $p_x=\dot{x}-y$, $p_y=\dot{y}+x$, and $p_z=\dot{z}$ are defined as pseudo momenta. 

The normal form is obtained by applying a symplectic linear change of variables, where parameters are found by studying the linearized terms of the CRTBP dynamics, in terms of its eigenvalues and eigenvectors. Following the procedure described in [\citenum{jorba1999dynamics}], after applying linear transformations, the Hamiltonian reaches the form 
\begin{equation}\label{H1}
H_2=\lambda_1xp_x+\frac{\omega_1}{2}\left(y^2+p_y^2\right)+\frac{\omega_2}{2}\left(y^2+p_y^2\right)-\sum_{n\geq2} c_n \rho^n P_n\left(\frac{x}{\rho}\right),
\end{equation}
where the constants are defined as 
\begin{equation}
\lambda_1^2=\frac{c_2-2-\sqrt{9c_2^2-8c_2}}{2}, \quad \omega_1^2=\frac{c_2-2+\sqrt{9c_2^2-8c_2}}{2}, \quad 
\omega_2^2=c_2.
\end{equation}
The evaluation of these constants comes from the study of the characteristic equation of the linearized system, where the $x$-$y$ motion characteristic equation has four roots, two real and two imaginary roots given by $(\pm \lambda_1, \pm  \omega_1)$, while the $z$ direction motion (decoupled) is characterized by two imaginary eigenvalues $\pm \omega_2$.

Finally, the Hamiltonian in normal form is obtained by applying a complex normal transformation:
\begin{subequations}
\begin{align}
x=q_1, \quad y=\frac{q_2+\sqrt{-1}p_2}{\sqrt{2}}, \quad z=\frac{q_3+\sqrt{-1}p_3}{\sqrt{2}},\\
p_x=p_1, \quad p_y=p_2, \quad p_z=\frac{\sqrt{-1}q_3+p_3}{\sqrt{2}}.
\end{align}
\end{subequations}
Using this new set of variables, the complex normal form Hamiltonian is given by 
\begin{equation}
H_3=\lambda_2 q_1 p_1 +\sqrt{-1}\omega_1 q_2 p_2 +\sqrt{-1} \omega_2 q_3p_3-\sum_{n\geq2} c_n \rho^n P_n\left(\frac{x}{\rho}\right),
\end{equation}
where $x$ and $\rho$ can be computed in terms of the new variables using the previous transformations. Hamiltonian $H_3$  decouples the variables in the linear leading term and it decouples of effects of the stable and unstable manifolds. The KOF uses the equations of motion evaluated from $H_3$ (obtained using Hamilton's equations) to perform the propagation of uncertainties, and subsequently it transforms variables back in the original set. However, every \textit{truth} model used for performance and comparison purposes has been embedded with the original set of ODE described in system \eqref{sys1}.

\subsection{Prediction of Central Moments}
\begin{figure}[h!]
	\centering
	\includegraphics[width=.65\textwidth]{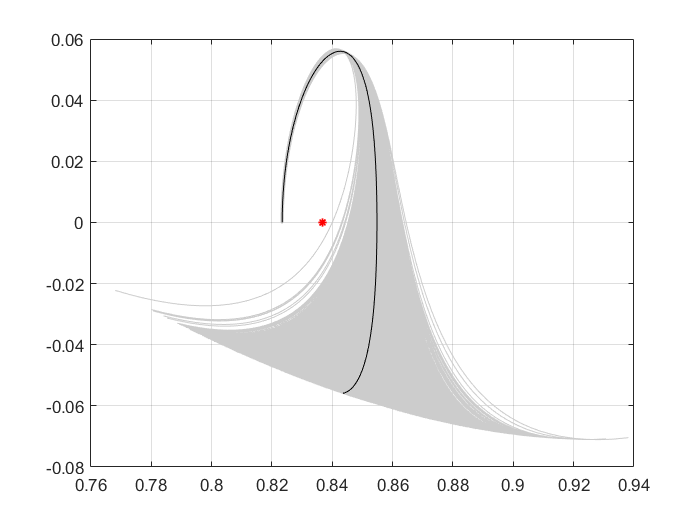}
	\caption{Monte Carlo propagation of $10^4$ orbits for $t_f = 2$ given the initial PDF $\mathcal N ({\bf{x}}_{0},{\bf P}_0)$. Projection on the $(x,y)$ plane.   }
	\label{fig:chaos}
\end{figure}

The correct propagation of uncertainties in time is offered on an $\mathcal L_1$ halo orbit application for the Earth-Moon system, characterized by $\mu = 0.012153281419431$ and $\gamma = 0.150944856782713$. The novel technique based on the KO adapts the spectral decomposition of the dynamics in terms of its eigenfunctions to accurately predict the state central moments at any given time (without performing any numerical propagation). Consider the halo orbit connected to the initial condition
\begin{align}
    {\bf{x}}_{0} = 
    \left[
    0.823376807050253 \quad 0  \quad 0.001386166961157 \quad 0 \quad 0.126366690232230 \quad 0
    \right]^T
\end{align}
and with given initial covariance matrix ${\bf P}_0 = \sigma_0^2 {\bf I}_{6\times 6} $, with $\sigma_0 = 0.0001 $. This initial distribution shows the exponential divergence nature of the CRTBP. Figure \ref{fig:chaos} portraits the evolution of $10^4$ orbits randomly chosen form the Gaussian PDF $\mathcal N ({\bf{x}}_{0},{\bf P}_0)$, up to the final time $t_f = 2$. 
The projection of these orbits into the $(x,y)$ plane illustrates the divergence of the states. The initial small covariance matrix rapidly increases and the figure shows the family of orbits form the conical shave connected to a chaotic system. Therefore, even before reaching half of a revolution, most of the orbits already strongly separate from the mean (periodic) orbit, in black. Thus, in just a short time span, small perturbations from the mean at the initial time lead to extremely large deviations, that make the prediction of uncertainties a challenging problem.   

\begin{figure}[h!]
	\centering
	\includegraphics[width=\textwidth]{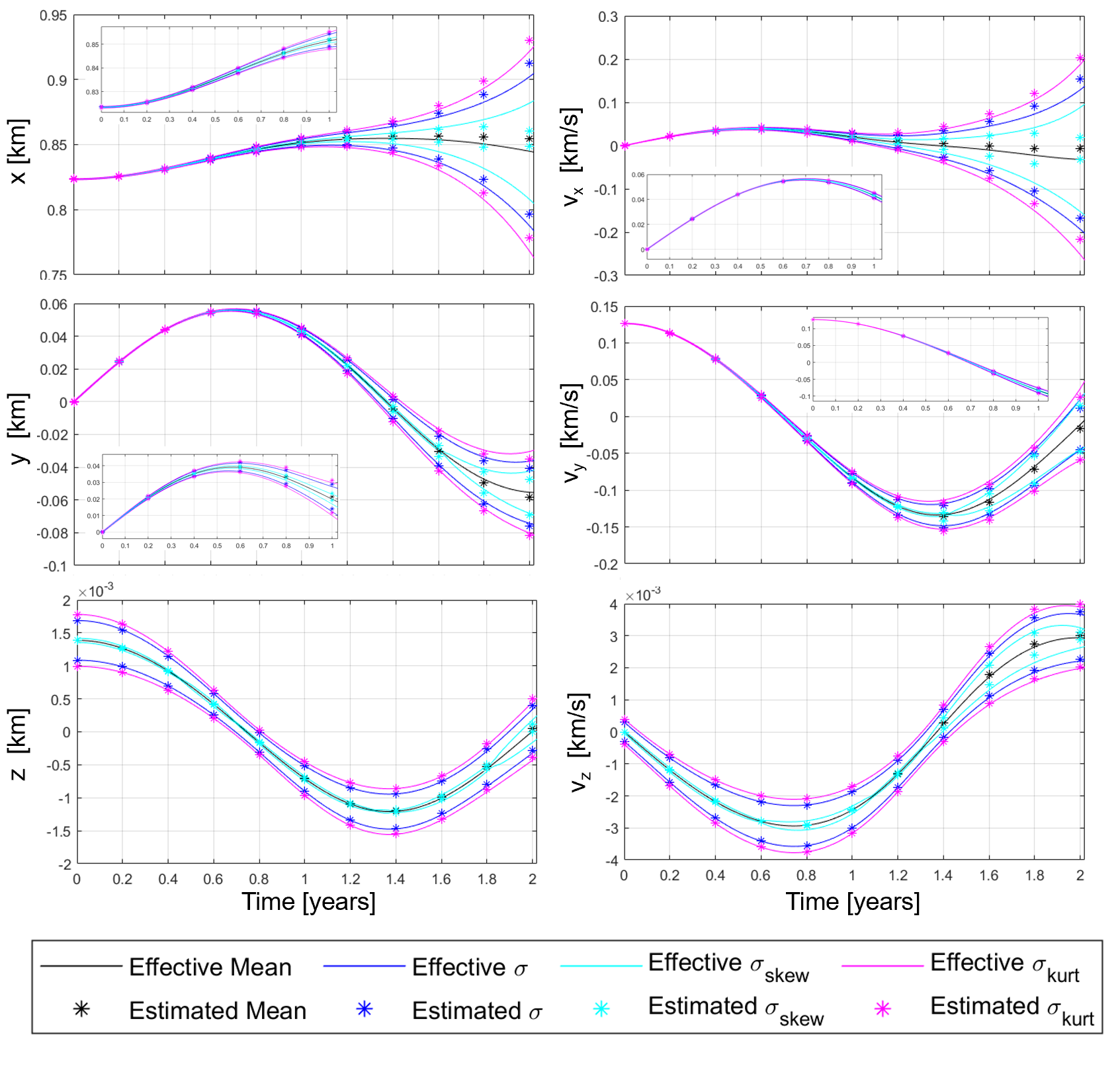}
	\caption{Prediction of raw moments of the state PDF for a Halo orbit orbiting around $\mathcal{L}_1$ in the earth-Moon system.   }
	\label{fig:mom}
\end{figure}

The newly developed technique has been applied to the CRTBP in the Earth-Moon system and results have been validated through a Monte Carlo analysis, where the predicted central moments of the state have been compared to the effective moments evaluated from all of the Monte Carlo runs. Figure \ref{fig:mom} shows the time evolution of the uncertainties represented with the state raw moments. Indeed, the continuous lines represents the true evolution of the central moments of the system: the mean in black, the covariance, as $3{\boldsymbol{ \sigma}}$ standard deviation, in blue, the skewness in cyan and the kurtosis in magenta. The values of skewness and kurtosis are evaluated similarly to the covariance of the system, with an analogy to the standard deviation:
\begin{align}
    \sigma_{j} = \left(\dfrac{\sum_{i=1}^{N} \left( x_{i,j} - \hat {x}_{j}\right)^2}{N-1} \right)^{1/2} \ \ \forall j = 1,\dots,n \label{eq:std}\\
    \sigma_{skew,j} = \left(\dfrac{\sum_{i=1}^{N} \left( x_{i,j} - \hat {x}_{j}\right)^3}{N-1} \right)^{1/3} \ \ \forall j = 1,\dots,n\\
    \sigma_{kurt,j} = \left(\dfrac{\sum_{i=1}^{N} \left( x_{i,j} - \hat {x}_{j}\right)^4}{N-1} \right)^{1/4} \ \ \forall j = 1,\dots,n
\end{align}
where ${N}$ is the number of Monte Carlo runs. Therefore, for each component of the state, the high order central moment is evaluated by analysing all of the $10^4$ simulations. On the contrary, the KO uncertainty propagation predicts the central moments of the state PDF every 0.2 years. The values of the predicted moments are reported, with the same colors used for the Monte Carlo, as points. The correct prediction of the mean and uncertainties of the system is asserted by the overlapping of the points over the continuous lines. The accuracy of the estimation of the moments decreases as the integrating time increases, meaning that, for extremely large propagation times, the prediction of the state PDF is unfeasible. The CRTBP is strongly diverging: thus, high order central moments after long time steps require an elevated amount of runs in order to have correct indicating values. Moreover, the figure shows the high level of accuracy in the first half of the simulation, for times shorter than 1 year, zoomed in in the figure. Station keeping applications, in filtering, usually receive measurements in the section where KO shows an extremely accurate prediction of moments. The KO technique can predict central moments up to any arbitrary high order. However, due to the fact that odd moments can be negative, the accuracy of the prediction of even moments is higher than their odd counterpart. Indeed, some values of the skewness, in the figure, are slightly off by the end of the simulation, where the state PDF largely expands without control. In the remaining sectors, the prediction of odd moments outperforms any other uncertainty prediction technique based on the Gaussian approximation of the state distribution (such as UT), since they provide a null value for any odd moment. 

\begin{figure}[h!]
	\centering
	\includegraphics[width=\textwidth]{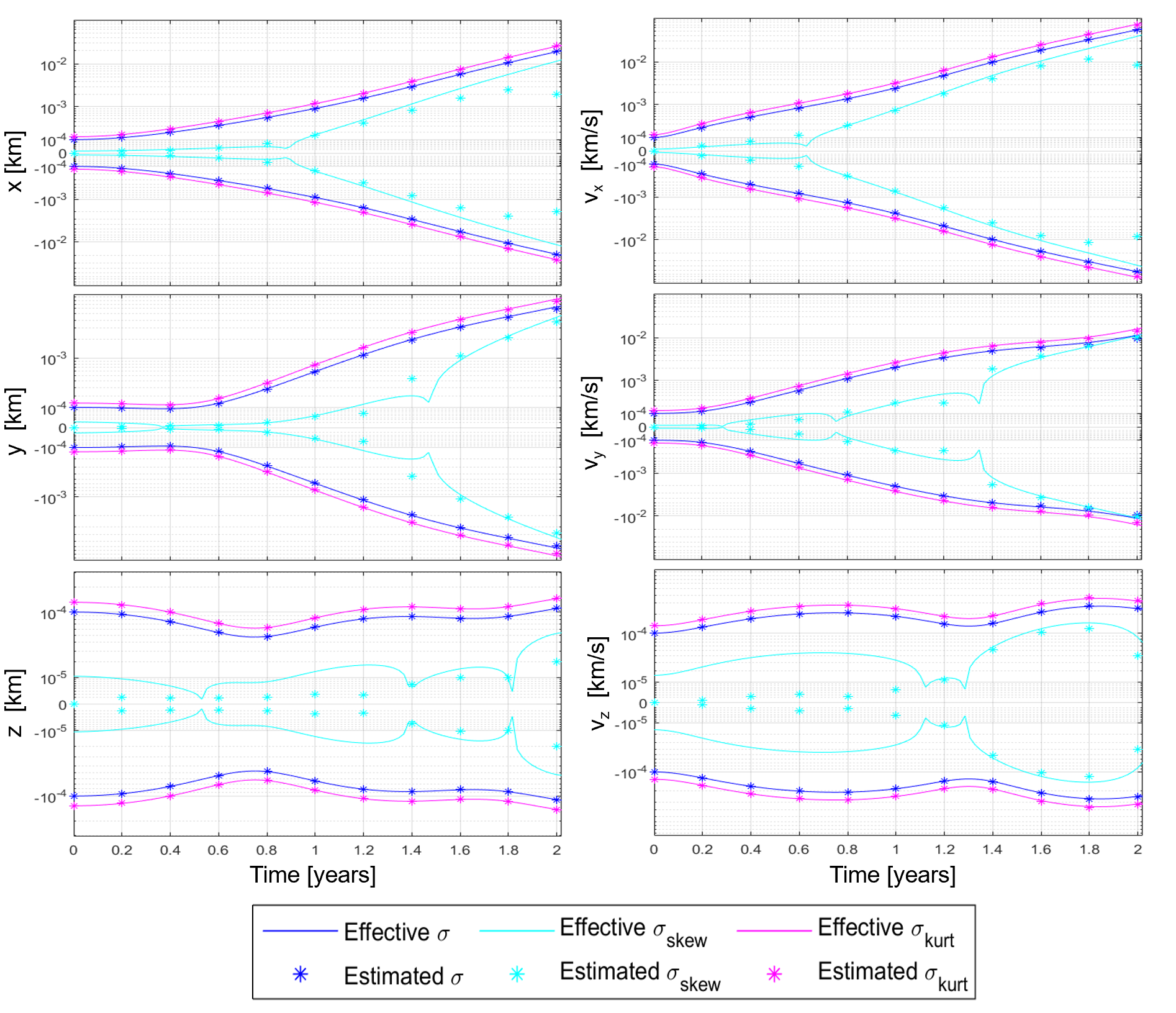}
	\caption{Prediction of central moments of the state PDF for a Halo orbit orbiting around $\mathcal{L}_1$ in the earth-Moon system.   }
	\label{fig:central}
\end{figure}

The central moments estimation can be appreciated by reporting the uncertainties as around their mean. Indeed, while Figure \ref{fig:mom} shows the correct prediction of the family of orbits, in terms of its overall possible trajectories, Figure \ref{fig:central} focuses on the spread around the mean. Therefore, Figure \ref{fig:central} displays similar results as Figure \ref{fig:mom}, but with a new prospective centered at the current mean. Figure \ref{fig:central} shows the chaotic nature of the problem, as the uncertainties, in terms of their $\sigma$ value, increase exponentially and the spread of all the possible resolutions of the state creates a conical shape. The KO prediction is able to keep track of the central moments of the system, especially during the first half of the simulation. Both the state covariance and kurtosis expand without any boundaries. The star points show the correct estimation performed by the KO technique. The new methodology is keeping an accurate prediction of the state distribution even for a long period of time, even if slightly reducing the accuracy. The figure has a logarithmic scale for the values of moments, highlighting the diverging nature of the CRTBP dynamics. Thus, it is worth noticing the exponential increase of the spread of the state PDF, with standard deviations that are two order of magnitude bigger after merely three quarters of a full revolution.

This application assesses the correct prediction of uncertainties and opens the road to the KOF, a complete filtering algorithm, that completes the propagation of mean and covariance with the knowledge obtained externally though noisy observations.  

\subsection{Orbit Determination}
While the previous problem focused on the corrected prediction of the moments of the state PDF under an highly nonlinear dynamics, a station keeping application is now proposed, where the system receives external information from a set of sensors, and the whole KOF algorithm is implemented. Therefore, let consider the Sun-Earth system, characterized by $\mu = 3.003410642560030e-06$ and $\gamma = 0.009970325504020$, and a Lyapunov orbit around the $\mathcal L_1$ libration point. The initial true state is selected randomly from a Gaussian distribution with initial state mean
\begin{align}
    {\bf{x}}_{0} = 
    \left[
    0.989826595322 \quad 0  \quad 0 \quad 0 \quad 0.00137295958289 \quad 0
    \right]^T
\end{align}
and covariance matrix ${\bf P}_0 = \sigma^2 {\bf I}_{6\times 6} $ with $\sigma = 10^{-4}$. The KOF scales and translates uncertainties and central moments to be consistent with the normal form representation of the Hamiltonian $H_3$. The propagation of the truth, used to evaluate the performance of the filter in terms of accuracy error, is propagated numerically using and high precision Runge-Kutta 7/8 integrator. Accuracy errors and estimates are then evaluated by scaling back the results obtained from the KOF, in order to agree with the classic representation of the dynamics. 

Nonlinear angle measurements, azimuth and elevation from the center of the Earth, are received by the filter every 0.4 year, which translates to about a total of 15 observations each 2 revolutions;
\begin{align}
    y_1 &= \arctan \left( \dfrac{x_2}{x_1 - 1 + \mu}\right) + \eta_1\\
    y_2 &= \arcsin \left( \dfrac{x_3}{\sqrt{(x_1 - 1 + \mu)^2 + x_2^2 + x_3^2}}\right) + \eta_2
\end{align}
where each $\eta$ is assumed Gaussian noise with zero mean and a standard deviation of 10 arcsec.

\begin{figure}[h!]
    \makebox[\textwidth][c]{\includegraphics[width=1.2\textwidth]{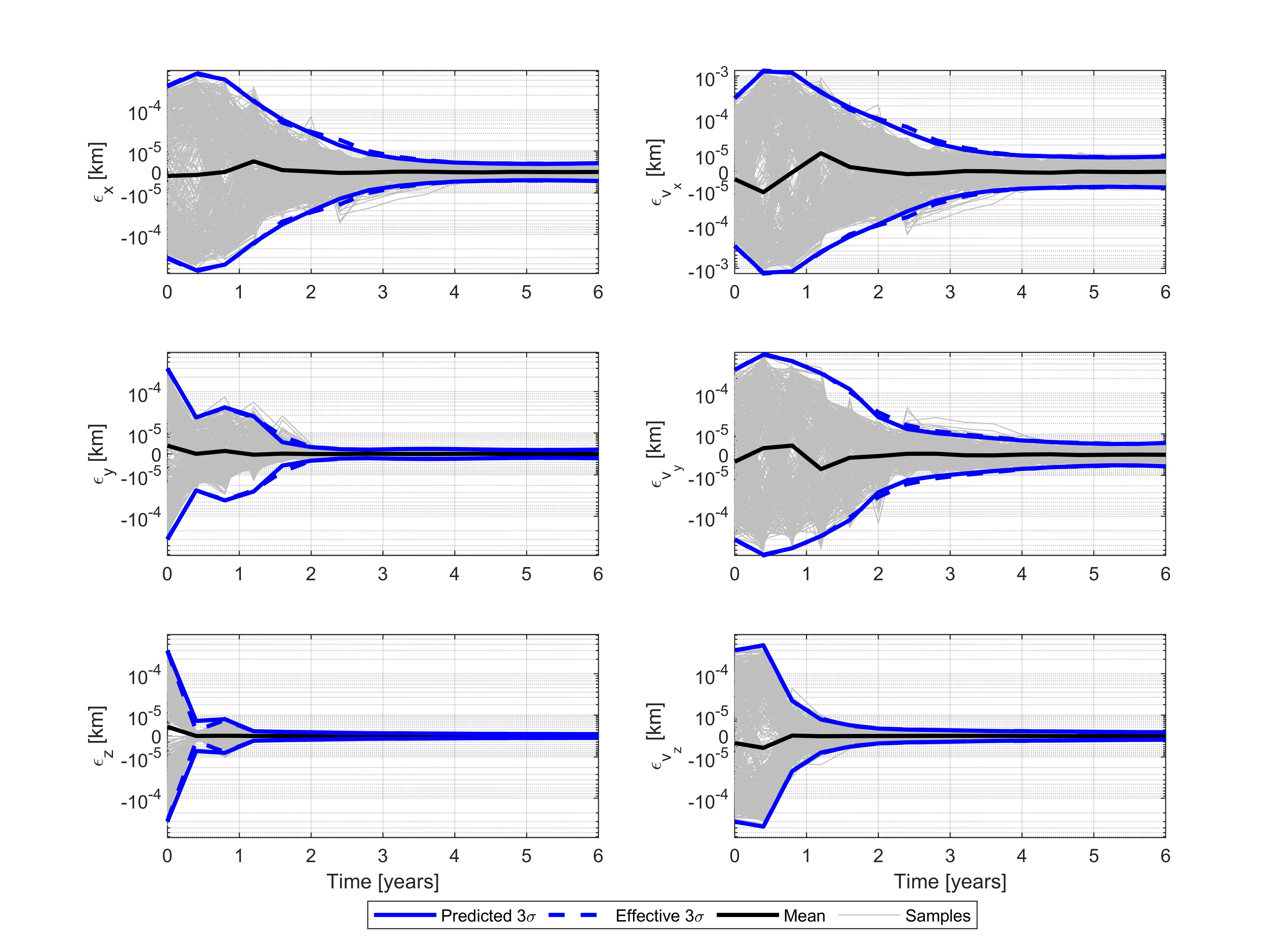}}
    \caption{500 runs Monte Carlo analysis of the KOF for a Lyapunov orbit around $\mathcal{L}_1$.  }
    \centering
    \label{fig:conv}
\end{figure}

A Monte Carlo analysis is performed to assess the performance of the filter. The initial level of uncertainties is so large that the accuracy level at steady state is two orders of magnitude smaller than the initial standard deviation. Therefore, in order to show convergence in a more pleasant manner, a logarithmic scale for the errors has been applied. Thus, Figure \ref{fig:conv} shows, for each component of the state, the 500 runs Monte Carlo analysis performed with the KOF on the selected Lyapunov orbit. Indeed, each gray line represents the errors connected to the $j$-th run of the Monte Carlo, for each $i$-th state component, evaluated as
\begin{align}
    \epsilon_{i,j} &= \hat x_{i,j} - x_{i,j}
\end{align}
The mean of the errors is portrayed with black lines, calculated as
\begin{align}
    \bar \epsilon_i &= \dfrac{1}{N}\sum_{j=1}^N \epsilon_{i,j}
\end{align}
where $N$ is the number of total Monte Carlo runs. The black lines settle on the zero value, proving that the KOF is an unbiased filter, as expected from the Minimum Mean Square Error (MMSE) principle on which the update of the filter is based on. The Monte Carlo analysis provides also information about the spread of the state error and the level of uncertainties. Figure \ref{fig:conv} has multiple pairs of dashed and continuous blue lines. Dashed blue lines represents the level of standard deviation, as $\pm 3 \sigma_{eff}$, obtained from the Monte Carlo runs and calculated using Equation \eqref{eq:std}: they represent the \textit{effective} level of accuracy of the filter. These dashed lines are calculated, for each time step, by considering the values of the state errors among all the 500 simulations and they represent the actual behaviour of the filter. On the contrary, the continuous blue lines report the \textit{estimated} level of uncertainties, again as $\pm 3 \sigma_{pred}$, predicted by the filter. Thus, for each time step, these lines are calculated extracting the error std directly from the update covariance of the filter, as $\sigma_{pred,i} = \sqrt{{\bf P}^+_{ii}}$. By looking at Figure \ref{fig:conv}, we can assess the consistency of the KOF, since the dashed and continuous blue lines overlap, meaning that the filter is able to correctly predict its own uncertainty. The KOF shows convergence and consistency, where steady state accuracy levels are reached rapidly after a few updates. The large initial uncertainties are rapidly reduced around the most current estimate, and the updated covariance of the filter correctly represents the spread of the state error PDF.

\begin{figure}[h!]
    \makebox[\textwidth][c]{\includegraphics[width=1.2\textwidth]{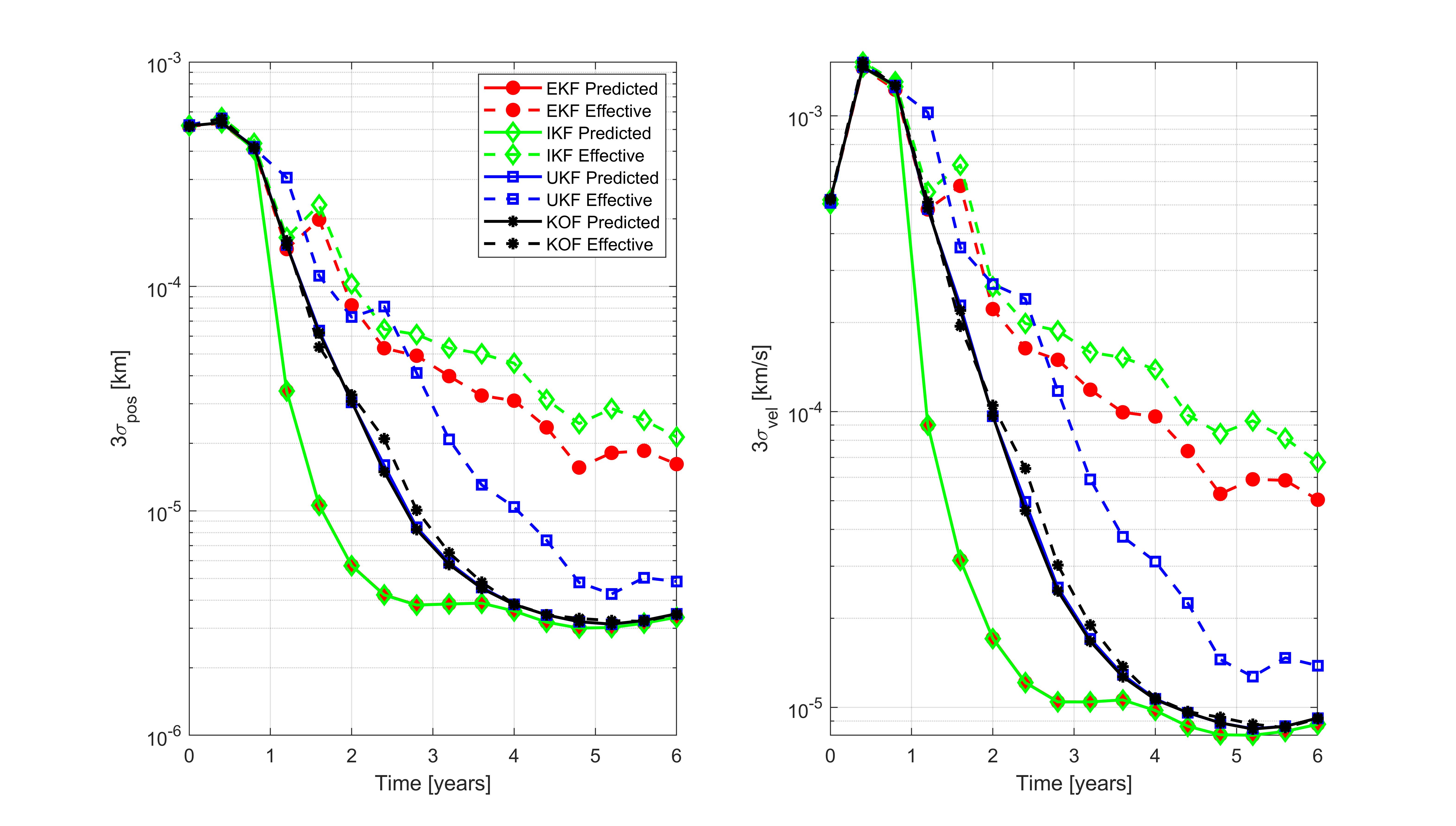}}
    \caption{Position and Velocity standard deviation analysis among EKF, IKF, UKF, and KOF.}
    \centering
    \label{fig:comp}
\end{figure}

The performance of the KOF can now be compared to other common filters: the EKF, the iterated extended Kalman filter (IKF), and the UKF have been selected as benchmark. The comparison in accuracy among different estimators is performed by analyzing the updated level of uncertainties, with a consistency check obtained from the Monte Carlo. Therefore, the 500 runs Monte Carlo analysis has been repeated for the EKF, IKF and UKF. Standard deviations for position and velocity are calculated and compared separately. The predicted stds are obtained directly from the trace of the updated covariance matrix, for each time step:
\begin{align}
    \sigma_{pos,pred} = \sqrt{{\bf P}^+_{11}+{\bf P}^+_{22}+{\bf P}^+_{33}}\\
    \sigma_{vel,pred} = \sqrt{{\bf P}^+_{44}+{\bf P}^+_{55}+{\bf P}^+_{66}}
\end{align}
while the effective stds levels, $\sigma_{pos,eff}$ and $\sigma_{vel,eff}$, are evaluated directly from the Monte Carlo runs. Thus, a consistent filtering algorithm has matching between $\sigma_{eff}$ and $\sigma_{pred}$. Figure \ref{fig:comp} reports the accuracy levels, in terms of error stds, of the four filters, displaying estimated values with continuous lines and the actual behaviour with dashed lines. The EKF is shown in red. The dashed red lines, both for position and velocity, are out of scale when compared to the continuous counterparts, which are overlapping with the green continuous lines. Indeed, the green lines are connected to the IKF. The EKF and the IKF share the same updated covariance matrix, while their estimate changes. Thus, the overlapping between the green and the red continuous lines matches with what is expected from theory: the two estimators approximate the posterior distribution of the state as Gaussian, with the same level of uncertainty (same stds), but different means. Indeed, while the EKF is a MMSE filter, the IKF is a filter based on the Maximum A Posteriori (MAP) principle, which outputs, as its estimate, the most likely state of the posterior distribution. However, the IFK and EKF share the same prediction step, based on the linearization of the dynamics, which is not sufficient to achieve an accurate approximation of the state prior distribution. Therefore, both filters are inconsistent and their dashed lines settle orders of magnitude above the continuous ones. The EKF and IKF believe that they are achieving higher accuracy than their actual results, and they are overconfident in their performance The difference between the green dashed lines and the red dashed lines is connected to the different update step performed by the two filtering techniques: a MMMSE one for the EKF, and a MAP one for the EKF. On the contrary, in blue, the UKF applies the unscented transformation in its prediction step to obtain a more accurate state prior distribution. Thus, the dashed blue line, which shows the effective accuracy level of the UKF, settles below the EKF and the IKF ones. However, the matching with the predicted uncertainty is missing and the filter is inconsistent as well. The UKF estimates a performance similar to the KOF, in black, shown by the overlapping between the black and the blue continuous lines. Therefore, the KOF is the only filter that converges with consistency. The KOF is able to correctly predict its own uncertainties and its estimate is the most accurate among the other filtering techniques. The dashed black lines are the lowest dashed lines for both position and velocity, and they overlap their continuous counterparts along the whole time-length of the simulation.


\section{Conclusions}
The propagation of uncertainties under highly nonlinear dynamics has been performed in the KO framework, where the solution of the system is approximated using a set of well defined orthogonal polynomials. Thanks to this representation, central moments of the state PDF can be predicted up to an arbitrary order. The new methodology has been paired with the Kalman measurement update to develop the KOF: a nonlinear filtering algorithm where both the state and measurement distributions are represented using the KO eigenfunctions. Numerical applications have assessed the robustness and accuracy of the new theory. Uncertainty prediction in the CRTBP has shown that the new technique is able to track the exponential spread of the spacecraft position and velocity PDF, providing estimates about the system's raw and central moments. Whenever information is provided in terms of external observations, the KOF is able to shrink the state uncertainties around the most current estimate. Monte Carlo analyses have assessed the accuracy, consistency, and robustness of the KOF, which has been proven to outperform other filtering techniques reported for comparison purposes.  


\section*{Acknowledgments}

The authors want to acknowledge the support of this work by the Air Force’s Office of Scientific Research under Contract Number FA9550-22-1-0092.

\bibliographystyle{AAS_publication}
\bibliography{libration_koopman}

\end{document}